\documentclass[a4paper,10pt]{mn2e}
\usepackage[utf8]{inputenc}
\usepackage{amsmath}
\usepackage{amssymb}
\usepackage{amstext}
\usepackage{graphicx}
\usepackage{deluxetable}
\usepackage{subcaption}
\usepackage{natbib}
\usepackage{float}

\newcommand{\mpy}{\, M_\odot \, {\rm yr^{-1}}}
\newcommand{\kms}{\rm \, km \, s^{-1}}
\newcommand{\cc}{\rm \, cm^{-3}}
\newcommand{\erg}{\rm \, erg \, s^{-1}}

\title{Observational Signatures of Galactic Winds Powered by Active Galactic Nuclei}
\author[Nims et al.] {Jesse Nims$^{1}$, Eliot Quataert$^{1}$, and Claude-Andr{\'e} Faucher-Gigu{\`e}re$^{2}$\\
$^{1}$Department of Astronomy and Theoretical Astrophysics Center, University of California, Berkeley, CA 94720-3411, USA \\
$^{2}$Department of Physics and Astronomy and CIERA, Northwestern University, 2145 Sheridan Road, Evanston, IL 60208, USA}

\begin{document}

\maketitle

\begin{abstract}

We predict the observational signatures of galaxy scale outflows powered by active galactic nuclei (AGN). Most of the emission is produced by the forward shock driven into the ambient interstellar medium (ISM) rather than by the reverse shock.    AGN powered galactic winds with energetics suggested by phenomenological feedback arguments should produce  spatially extended $\sim 1-10$ keV X-ray emission $\sim 10^{41-44} \erg$, significantly in excess of the spatially extended X-ray emission associated with normal star forming galaxies.   The presence of such emission is a direct test of whether  AGN outflows significantly interact with the ISM of their host galaxy.    We further show that even radio quiet quasars should have a radio luminosity comparable to or in excess of the far infrared-radio correlation of normal star forming galaxies.   This radio emission directly constrains the total kinetic energy flux in AGN-powered galactic winds.
Radio emission from AGN wind shocks can also explain the recently highlighted correlations between radio luminosity and the kinematics of AGN narrow-line regions in radio quiet quasars.  
\end{abstract}


\section{Introduction}

Galaxy scale outflows driven by active galactic nuclei (AGN) have been invoked to explain the close connection between  the properties of black holes and their host galaxies (e.g., the $M_{BH}-\sigma$ relation; \citealt{1998A&A...331L...1S}) and  the quenching of star formation in massive galaxies \citep{Springel2005}.    However, observationally identifying galactic outflows driven by AGN rather than star formation is non-trivial.   There has been significant progress in this direction in the last $\sim 5$ years, in particular through observations of nearby ultraluminous infrared galaxies and quasars (e.g., \citealt{2010A&A...518L.155F, Rupke2013,Cicone2014}) and the Type 2 quasar sample from SDSS (e.g., \citealt{Liu2013,2014MNRAS.442..784Z,2014ApJ...788...54G}).\par
The goal of the present paper is to predict several observational diagnostics of AGN-powered galactic winds.
Throughout we focus on simple, but robust, analytic estimates.     We also compare the resulting predictions  with analogous predictions for star formation powered winds in order to identify ways of disentangling these two energy sources powering galactic winds.   Our study builds on earlier work with related goals (e.g., \citealt{Jiang2011,2012MNRAS.420.1347F,Bourne2013}).

Throughout this paper, we utilize \citet{2012MNRAS.425..605F}'s (hereafter FGQ12) analytic model for energy conserving galactic outflows driven by AGN.   In particular, we consider galactic winds with kinetic luminosities $\sim 0.01-0.05 \, L_{\rm AGN}$ and  momentum fluxes of $\sim 1-10 \, L_{\rm AGN}/c$. These values are motivated by observed galactic winds attributed (at least in part) to AGN (e.g., \citealt{2012MNRAS.420.1347F,Arav2013,Cicone2014}) and by models of the $M_{BH}-\sigma$ relation (e.g., \citealt{King2003,DiMatteo2005,2012MNRAS.420.2221D}).

 In \S \ref{sec:setup} we  briefly review the physics of galactic winds powered by AGN, highlighting that the observed emission is likely to be dominated by the forward shock driven into the interstellar medium, rather than the reverse shock in the AGN wind (see Fig. \ref{fig:1}).    This leads to more luminous emission associated with AGN-powered galactic winds than predicted previously by FGQ12, who focused on the shocked wind emission because it is the latter that is critical for the dynamics of the outflow as a whole.    In \S \ref{sec:thermal} we present analytic and numerical estimates of the thermal X-ray emission of the shocked interstellar medium (ISM) produced by a combination of free-free and inverse Compton emission; we also compare this to  emission from the shocked wind.   In \S \ref{sec:nonthermal} we calculate the  non-thermal  emission from shock accelerated electrons and protons and the resulting radio to $\gamma$-ray luminosities.    We summarize our results and predictions in \S \ref{sec:discussion} and compare them to current observational constraints.

\section{Problem setup}\label{sec:setup}
\subsection{Energy conserving outflows}

\begin{figure}

 \includegraphics[bb=0 0 315 237,scale=.25,keepaspectratio=true,clip=true,trim=2pt 1pt 1pt 1pt]{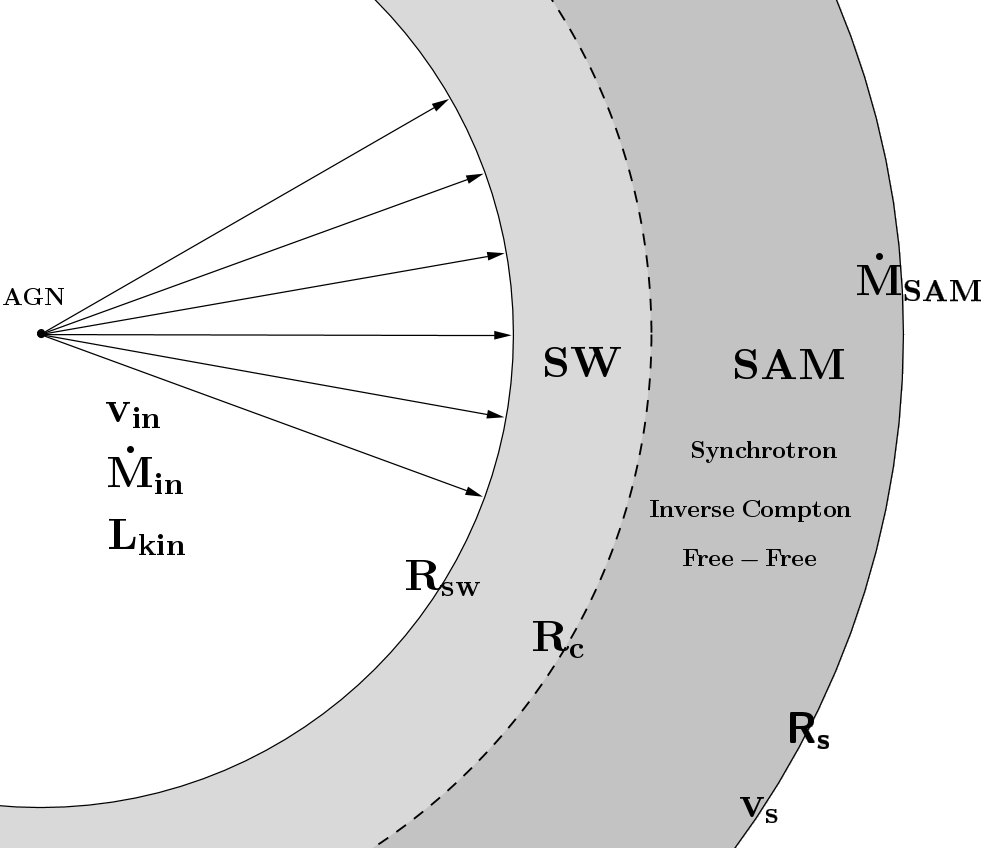}

 \caption{An AGN drives a wind with velocity $v_{in}$, mass loss rate $\dot M_{in}$, and kinetic luminosity $L_{kin}$.  This creates a forward shock with radius $R_s$ and velocity $v_s$ that propagates through the ambient medium and a reverse shock with radius $R_{sw}$. The two shocks are separated by a contact discontinuity at $R_c$.  Behind the forward shock is the shocked ambient medium (SAM), which will radiate through thermal free-free and inverse Compton emission as well as non-thermal synchrotron, inverse Compton and pion decay emission.  $\dot M_{\rm SAM}$ is the rate at which the forward shock sweeps up the ambient medium.  This is typically $\gg \dot M_{in}$.  If energy is conserved, the dynamics of the galactic-scale wind driven by the AGN (e.g., $R_{s}$ vs. time) depends solely on $L_{kin}$ and the density profile of the ambient medium (parameterized by eq. \ref{eq:densityfalloff}).           Note: this Figure is not to scale. For an 
 energy conserving outflow the shocked wind (SW) fills most of the volume. When it cools the SAM is much thinner than pictured.}
 \label{fig:1}
\end{figure}

\indent We use the energy conserving outflow model of FGQ12.   Figure \ref{fig:1} and Table 1 provide a schematic representation of the setup and define some of the key parameters we will use in this paper.   We assume that an AGN with luminosity $L_{\rm{AGN}}$ drives a wind from the accretion disk close to the black hole.    The interaction between the AGN wind and the surrounding ISM is determined solely by the density distribution of surrounding gas and the total kinetic power in the AGN wind, $L_{kin}$.\footnote{This
is only true when energy is conserved. In general, the magnitude of $v_{in}$ impacts the radiative losses and thus whether the assumption of energy conservation is reasonable.}  The latter can  be expressed in terms of the initial speed $v_{in}$ and mass loss rate $\dot M_{in}$ of the wind or the input momentum flux  $\dot{M}_{in} v_{in} = \tau_{in} (L_{\rm AGN}/c)$, where $\tau_{in}$ quantifies the initial wind momentum flux  relative to the  momentum flux in the AGN radiation field.

\begin{table}
\begin{center}
\caption{Fiducial Parameters Utilized in Analytic Estimates}
\begin{tabular}{|l|c|r|}
\hline
Parameter & Fiducial Value & Units\\
\hline
\(\displaystyle L_{\rm{AGN}}\)\tablenotemark{1} & \(10^{46}\) & erg \(\rm{s^{-1}}\)\\

\(L_{kin}\)\tablenotemark{2} & 0.05 & \( \displaystyle L_{\rm{AGN}}\)\\
\(\displaystyle n_{H,0} \)\tablenotemark{3} & 10 & \( \rm{cm^{-3}}\)\\
\(\alpha\)\tablenotemark{3} & 1 & -- \\
\(R_0\)\tablenotemark{3} & 100 & pc\\
\(v_{in}\)\tablenotemark{4} & 0.1 & c\\
\hline
\end{tabular}
\tablenotetext{1}{Bolometric luminosity of the AGN.}

\tablenotetext{2}{
Kinetic energy flux supplied by the AGN wind at small radii (see Fig. \ref{fig:1}).   Approximately half of this energy goes into the forward shock driven into the ambient medium.  The other half is thermal energy of the shocked wind bubble.}

\tablenotetext{3}{Parameters in the ambient medium density profile; see eq. (\ref{eq:densityfalloff}).}
\tablenotetext{4}{Input speed of the AGN wind on small scales (Fig. \ref{fig:1}).}
\end{center}
\end{table}
\

We  assume that the ambient medium has a density that decreases as a power law  with distance from the galactic nucleus:
\begin{equation}\label{eq:densityfalloff}
 n_H(R)=n_{H,0}\biggl(\frac{R_s}{R_0}\biggr)^{-\alpha}
\end{equation}
where $n_{H,0}$ is the number density of the ambient medium at an arbitrary chosen fiducial radius $R_0$.
We model both the AGN wind and the ambient medium as spherically symmetric and we neglect the realistic multiphase structure of the ISM of galaxies.    These are clearly gross over-simplifications.    We discuss this further in \S \ref{sec:discussion}.

When the wind sweeps up its own mass in
the ambient ISM, two shocks are formed that are separated by a contact discontinuity (Fig. \ref{fig:1}).   The innermost shock is a reverse shock with radius $R_{sw}$ that acts on the wind from the AGN;
for this reason, it is often called the wind shock. The outer shock with radius $R_s$ and velocity $v_{s}$ propagates outward through the ambient medium.  

The total input kinetic power is
\begin{equation}\label{eq:Lkin}
 L_{kin}=\frac{1}{2}\dot{M}_{in}v_{in}^2 \approx \left(1 + \frac{2}{3} \, \frac{\alpha - 2}{3-\alpha}\right) \dot{M}_{\rm{SAM}} v_s^2
\end{equation}
where $\dot{M}_{\rm{SAM}}=4\pi n_H(R_s)m_p R_s^2 v_s$ is the mass of the ambient medium swept up per unit time by the forward shock (this is in general $\gg \dot M_{in}$).   The second equality in equation (\ref{eq:Lkin}) follows from the fact that approximately half of the energy supplied by the AGN goes into energy of the forward shock; the other half is in the thermal energy of the shocked wind.
In the energy conserving case assumed in this paper $L_{kin}$ is independent of shock radius.

FGQ12  estimated the emission produced by the shocked wind (SW) as a probe of the power carried by AGN outflows.   They also showed that under many circumstances the SW does not
cool efficiently, making AGN powered galactic winds essentially energy conserving.  
The key reasons for this are that the wind shock produces very high temperature gas with
\begin{equation}
 T_{SW}=2.0\times 10^{10}\; \rm{K} \; \times \; \biggl(\frac{v_{in}}{0.1 c} \biggr)^2
 \label{eq:T-sw}
\end{equation}
and that the shocked wind density declines as $1/R^2$ which leads to shocked wind densities substantially lower than the SAM for most parameters of interest.

By contrast, as the AGN wind sweeps up the ambient medium the forward shock decelerates producing cooler gas with
\begin{equation}\label{eq:SAMtemp}
 T_{SAM}=\frac{3\mu}{16k}m_pv_s^2=2.3\times 10^{7}\; \rm{K} \biggl(\frac{v_s}{1000\;\rm{km\; s^{-1}}}\biggr)^2.
\end{equation}
More accurately, $T_{SAM}$ is  the temperature of the shocked protons. The electrons will
be heated to a somewhat lower temperature at the shock.   For the shocked wind, FGQ12 showed that this distinction between proton and electron temperatures is important and modifies the cooling rate.   
Two-temperature effects are less important for the SAM, however, because of the lower temperatures and higher densities.   
As a result, $T_e \approx T_{SAM}$.   
The exception is low ambient ISM densities and small radii, in which case the forward shock speed is not much smaller than $v_{in}$.   

The numerical results given in \S \ref{sec:numerics} include the effects of 
finite electron-proton equilibrium times following FGQ12 (see also \S \ref{sec:SW}) but since this is only important at very small radii we do not account for this in our analytic estimates that follow.

In \S \ref{sec:SW} we show that the denser SAM is typically more luminous than the SW.  We thus focus primarily on predictions of the emission from the SAM.       

The mean ISM densities of galaxies hosting powerful AGN can be large, $\sim 10^{3-4} \,\rm{cm^{-3}}$ on scales of $\sim 100$ pc if observations of nearby ULIRGs are representative (\citealt{1998ApJ...507..615D}).   These densities would be sufficient to quench an AGN powered galactic wind prior to its reaching  radii $\gtrsim$ kpc (FGQ12).    It is likely, however, that AGN winds follow paths of least resistance and interact with gas having densities significantly less than the formal volume averaged gas density.  We thus choose  characteristic ambient densities that produce galaxy-scale winds with properties broadly suggested by observations:  velocities of $\sim 1000 \, \rm{km \, s^{-1}}$ at radii $\sim 0.1-1$ kpc (e.g., \citealt{2010A&A...518L.155F,Cicone2014}). Momentum conservation implies that the mean density swept up by the AGN wind must then be
\begin{multline}
 \bar{n}_H(<R_s)\approx 50\;\rm{cm^{-3}} \tau_p \, \biggl(\frac{L_{\rm{AGN}}}{10^{46}\; \rm{erg\;s^{-1}}}\biggr) \\ \times \biggl(\frac{R_s}{100\;\rm{pc}}\biggr)^{-2} \biggl(\frac{v_s}{1000\; \rm{km\; s^{-1}}}\biggr)^{-2}
 \label{eq:n-constraint}
\end{multline}
where $\tau_{\rm{P}}=cP/(L_{\rm{AGN}}t_{flow})$ quantifies the momentum of the outflow $P$ at large radii; $\tau_p$ can be larger than the input momentum parameterized by $\tau_{in}$ because of work done during the energy conserving phase.

Energy conservation implies that the forward shock radius as a function of time in spherical symmetry is (see FGQ12)
\begin{multline}
\label{eq:Rs}
 R_s \approx 2 \;\rm{kpc} \, \biggl(\frac{3-\alpha}{2}\biggr)^{1/(5-\alpha)}\biggl(\frac{5-\alpha}{4}\biggr)^{2/(5-\alpha)}\\ \times \; \biggl(\frac{L_{kin}}{5 \times 10^{44}\; \rm{erg\;s^{-1}}}\biggr)^{1/(5-\alpha)} \biggl(\frac{n_{H,0}}{10\; \rm{cm^{-3}}}\biggr)^{-1/(5-\alpha)} \\ \times \;\biggl(\frac{R_0}{100\; \rm{pc}}\biggr)^{-\alpha/(5-\alpha)}\biggl(\frac{t}{10^6 \;\rm{yr}}\biggr)^{3/(5-\alpha)}
\end{multline}
and the forward shock velocity is given by\footnote{The factor of 3 was incorrectly placed outside the large brackets in eq. A5 of FGQ12. This is the corrected formula.  This typographical error did not propagate to other equations in Appendix A of FGQ12.}
\begin{multline}
\label{eq:vs}
 v_s=\frac{1}{2}\biggl [6\frac{(3-\alpha)L_{kin}}{(5-\alpha)\pi R_0^\alpha n_{H,0} m_p} \biggr ]^{1/3} R_s^{(\alpha-2)/3} \\
\approx  3400\; \rm{km\;s^{-1}}\biggl (\frac{6-2\alpha}{5-\alpha}\biggr)^{1/3} \biggl(\frac{L_{kin}}{5 \times 10^{44}\; \rm{erg\;s^{-1}}}\biggr)^{1/3} \\ 
 \times \; \biggl(\frac{R_0}{100\; \rm{pc}}\biggr)^{-\alpha/3}\biggl(\frac{n_{H,0}}{10\; \rm{cm^{-3}}}\biggr)^{-1/3}  \biggl(\frac{R_s}{100\; \rm{pc}}\biggr)^{(-2+\alpha)/3}.
\end{multline}

Note that for $\alpha = 1$, which we will take as our fiducial value in what follows, $R_s \propto t^{3/4}$, $v_s \propto R_s^{-1/3}$, and $T_{SAM} \propto R_s^{-2/3}$.   
Moreover, for a typical quasar lifetime of $\sim 10^7$ yr (e.g., \citealt{Martini2004}), equations (\ref{eq:Rs}) \& (\ref{eq:vs}) shows that the forward shock can readily reach distances of $\gtrsim 1$ kpc with velocities $\sim 1000 \kms$.

The timescale on which the volume of the SAM changes substantially is the flow timescale
\begin{multline}\label{eq:tflowScaled}
t_{flow} =\frac{R_s}{v_s}\approx 2.8\times 10^{4}\; \rm{yr} \, \biggl(\frac{6-2\alpha}{5-\alpha}\biggr)^{1/3} \biggl(\frac{n_{H,0}}{10\; \rm{cm^{-3}}}\biggr)^{1/3}\\ 
\times \; \biggl(\frac{L_{kin}}{5 \times 10^{44}\; \rm{erg\;s^{-1}}}\biggr)^{-1/3} \biggl(\frac{R_0}{100\; \rm{pc}}\biggr)^{\alpha/3} \biggl(\frac{R_s}{100\; \rm{pc}}\biggr)^{(5-\alpha)/3}.
\end{multline}
The free-free luminosity produced by the SAM
depends on  whether the cooling time  is short or long compared to this flow time. As we shall see, the IC luminosity depends only weakly on whether the SAM cools.

\section{Thermal Emission from the SAM}\label{sec:thermal}
In this section we determine the importance of both free-free and Compton cooling of the SAM.   We present  results for the emission produced by the SAM as a function of the forward shock radius $R_s$ and the parameters of the input AGN wind and ambient density profile.  In our analytic estimates, we ignore relativistic effects to increase the clarity of the presentation. We relax this assumption in our  numerical results in \S 3.4.    In both our analytic and numerical calculations we assume that at a given radius $R_s$, the emission from the SAM is dominated by matter newly shocked at that radius, rather than material shocked at a smaller radius that adiabatically expands as the shock moves outwards.  For the ambient medium density profiles we consider (with $0.5 \lesssim \alpha \lesssim 2$), this is a good approximation.    

\subsection{Inverse Compton emission}
\label{sec:IC}

Using equation (\ref{eq:densityfalloff}) the column density 
of the swept up SAM when the shock is at a radius $R_s$ is given approximately by
\begin{equation}
 N_{SAM} \approx \frac{\int_0^{R_s} n_H(R)R^2\rm{d}R}{R_s^2}
\end{equation}
The total luminosity from photons scattered by the swept up SAM is then
\begin{equation}
 L_{IC}\approx  L_{AGN} \sigma_T N_{SAM} \left(1+\frac{4}{3}<\gamma^2 \beta^2>\right)
 \label{eq:IC}
\end{equation}
Equation (\ref{eq:IC}) has two parts:  the first term in ( ) represents the scattered AGN radiation while the second term ($\propto \, <\gamma^2 \beta^2>$) accounts for the additional energy (of the shocked SAM) imparted to the photons during scattering.   The latter is $\lesssim 1$ for the SAM at all shock radii of interest and so the IC emission from the SAM is dominated by the pure Thomson scattering contribution.
 
Given this, the SAM IC contribution directly probes $\dot M_{SAM}$, the galaxy-scale mass loss rate driven by the AGN:
\begin{equation}
L_{IC} \approx \frac{L_{AGN} \sigma_T \dot M_{SAM}}{4 \pi R_s v_{s}}.
\end{equation}
Scaling to fiducial parameters we can also express the SAM IC emission as
\begin{multline}
 L_{IC}\approx 10^{43} \;\rm{erg\; s^{-1}} \biggl(\frac{2}{3-\alpha}\biggr) \biggl(\frac{L_{\rm{AGN}}}{10^{46} \; \rm{erg\;s^{-1}}}\biggr) \\ \times \;  \biggl(\frac{n_{H,0}}{10\; \rm{cm}^{-3}}\biggr)\biggl(\frac{R_0}{100\; \rm{pc}} \biggr)^{\alpha}\biggl(\frac{R_s}{100\; \rm{pc}} \biggr)^{1-\alpha}
\end{multline}
To calculate the  spectrum of the scattered radiation, we use the AGN spectrum from \cite{2004MNRAS.347..144S}:
\begin{displaymath}
 L_E(E_{ph})= \left\{
  \begin{array}{lr}
   A*E_{ph}^{-0.8}e^{-E_{ph}/(200 {\rm keV})} &  E_{ph}>2 \, {\rm keV} \\
   B*E_{ph}^{-1.7}e^{-E_{ph}/(2 {\rm keV})} &  10 \, {\rm eV} <E_{ph}<2 \, {\rm keV}\\
   C*E_{ph}^{-0.6}               &  1 \, {\rm eV} <E_{ph} < 10 \, {\rm eV}
   \label{eq:imputspectrum}
  \end{array}
\right.
\end{displaymath}
where A, B, and C are normalization constants which must be set to ensure continuity of the spectrum. 
Because $\gamma \beta \lesssim 1$, the output spectrum of the radiation scattered by the SAM will match the AGN spectrum quite closely.

The timescale for cooling of the SAM via inverse Compton scattering depends on the energy loss rate of the electrons in the SAM.   This is set  by the second term in ( ) in equation (\ref{eq:IC}), not the first.

The timescale for the newly shocked SAM to cool via inverse Compton scattering is thus
\begin{multline}
t_{IC}=\frac{3n_{SAM}kT}{2\epsilon_{IC}} \approx 4 \times10^5\; \rm{yr}  \biggl(\frac{R_s}{100\; \rm{pc}}\biggr)^2\biggl(\frac{10^{46}\; \rm{erg\;s^{-1}}}{L_{\rm{AGN}}}\biggr) 
\end{multline}
where $\epsilon_{IC}$ is the IC emissivity per unit volume (eg, \citealt{1979rpa..book.....R}) and we have assumed non-relativistic temperatures, as is appropriate for the SAM.   
The corresponding  rate at which the  electrons lose energy to IC cooling is
\begin{equation}\label{eq:LcompNoCool}
 \dot{E}^{SAM}_{e,IC}=\frac{t_{flow}}{t_{IC}}L_{kin}.
\end{equation}
\subsection{Free-Free Emission}
\label{sec:ff}
Using the energy conserving shock solution, the timescale for the SAM to cool through free-free emission is
\begin{multline}\label{eq:tff}
 t_{ff}=\frac{3n_{SAM}kT}{2\epsilon_{ff}} \approx 2 \times10^{6}\; \rm{yr} \biggl (\frac{6-2\alpha}{5-\alpha}\biggr)^{1/3} \biggl(\frac{R_s}{100\; \rm{pc}}\biggr)^{(4\alpha-2)/3} \\ 
 \times \; \biggl(\frac{R_0}{100\; \rm{pc}} \biggr)^{-4\alpha/3} \biggl(\frac{L_{kin}}{5 \times 10^{44}\; \rm{erg\;s}^{-1}}\biggr)^{1/3}
 \biggl(\frac{n_{H,0}}{10\; \rm{cm}^{-3}}\biggr)^{-4/3}
\end{multline}
where $\epsilon_{ff}$ is the free-free emissivity per unit volume (eg, \citealt{1979rpa..book.....R}).
 The total free-free emission when $t_{ff}>t_{flow}$ is
\begin{equation}\label{eq:LffNoCool}
 L_{ff}=\frac{t_{flow}}{t_{ff}}L_{kin}
\end{equation}
Using equations (\ref{eq:Lkin}), (\ref{eq:tflowScaled}) and (\ref{eq:tff}) this becomes
\begin{multline}
\label{eq:Lffnocool}
 L_{ff}=2 \times 10^{42}\;\rm{erg\;s^{-1}}  \biggl (\frac{6-2\alpha}{5-\alpha}\biggr)^{-2/3}\biggl(\frac{n_{H,0}}{10\; \rm{cm}^{-3}}\biggr)^{5/3}\\ \times \; \biggl(\frac{L_{kin}}{5\times 10^{44}\; 
 \rm{erg\;s^{-1}}}\biggr)^{1/3}\biggl(\frac{R_0}{100\; \rm{pc}}\biggr)^{5\alpha/3}  \biggl(\frac{R_s}{100\; \rm{pc}}\biggr)^{(7-5\alpha)/3} 
\end{multline}
We have set the frequency averaged gaunt factor to 1.2.

We use the standard non-relativistic  free-free emission results from  \citet{1979rpa..book.....R} to determine the spectrum.   The X-ray emission is of particular interest given the high shock velocities (eq. \ref{eq:vs}).    It is straightforward to show that so long as the SAM does not cool on a flow timescale, the free-free emission in the X-ray band produced by the SAM is given by (for $v_s \gtrsim 650 \; \rm{km\; s^{-1}}$)
\begin{multline}
\label{eq:LX}
 \nu L^{ff}_{\nu}(1\; \rm{keV})\simeq 2 \times10^{41} \; \rm{erg \;s^{-1}} \biggl (\frac{6-2\alpha}{5-\alpha}\biggr)^{-1} \times \\ \biggl(\frac{n_{H,0}}{10\; \rm{cm}^{-3}}\biggr)^{2} 
 \biggl(\frac{R_0}{100\; \rm{pc}}\biggr)^{2\alpha}  \biggl(\frac{R_s}{100\; \rm{pc}}\biggr)^{3-2\alpha} \biggl(\frac{v_s}{1000 \; \rm{km \; s^{-2}}}\biggr)^{-1} 
\end{multline}
where the limitation on shock velocity originates from the assumption that $h\nu \lesssim kT$, which allows us to eliminate the exponential term in the standard free-free emissivity.   Note that equation (\ref{eq:vs}) shows that the restriction $v_s \gtrsim 650 \kms$ is valid over a wide range of the parameters relevant to AGN-powered galactic winds, so that equation (\ref{eq:LX}) provides a reasonable analytic estimate of the expected X-ray emission.

\subsection{The Rapid Cooling Limit}
To determine whether the SAM cools on the flow timescale we define the net cooling time as
\begin{equation}
 t_{cool}=\frac{t_{IC}\,t_{ff}}{t_{ff}+t_{IC}}.
\end{equation}
When $t_{flow}>t_{cool}$ the SAM radiates the majority of its energy quickly and cools to a temperature $\ll T_{SAM}$ predicted by equation \ref{eq:SAMtemp}.   
In this rapid cooling limit, we cannot use equations (\ref{eq:LcompNoCool}) or (\ref{eq:LffNoCool})
because the energy loss rate of the electrons would exceed $L_{kin}$, violating energy conservation.
Instead, when $t_{flow}>t_{cool}$ all of the thermal energy
that is input into the SAM is radiated on a flow timescale.   The majority of the thermal emission still comes from gas at $\sim T_{SAM}$ even though the bulk of the SAM cools to a much lower temperature.   Thus in the rapid cooling limit\footnote{The equalities of this section are approximate because some of $L_{kin}$ goes to bulk kinetic energy of the SAM, rather than to thermal energy.}
\begin{equation}
 L_{ff}+\dot{E}^{SAM}_{e,IC}\approx L_{kin}
\end{equation}
and
\begin{equation}\label{eq:Lcomp}
  L_{ff}\approx \frac{t_{cool}}{t_{ff}}L_{kin}. 
\end{equation}
To smoothly interpolate between $t_{cool}<t_{flow}$ and $t_{cool}>t_{flow}$ we use
\begin{equation}
 L_{ff}\approx \max\biggl(1,\frac{t_{flow}}{t_{cool}}\biggr)\frac{t_{cool}}{t_{ff}}L_{kin}.
\end{equation}
Note that the possibility that the SAM radiates a significant fraction of its thermal energy is not in conflict with our assumption of an energy conserving galactic wind. The overall dynamics of the outflow are determined by the shocked wind; whether the SAM cools has only a small impact on the outflow solutions (e.g.,  \citealt{Koo1992}).

\subsection{Comparison to Shocked Wind Emission}
\label{sec:SW}

Here we determine the emission produced by the SW and show that it is small compared to the emission produced by the SAM derived in the previous sections.  We  utilize FGQ12's work on the thermodynamics of the SW.

The density of gas after passing through the wind shock at radius $R_{sw}$ (see Fig \ref{fig:1}) is given by
\begin{multline}
 n_{sw}(R_{sw})=\frac{\dot{M}_{in}}{\pi R_{sw}^2 m_p v_{in}} \approx \frac{\dot{M}_{in}}{\pi R_s^2 v_s m_p}\\ \approx 0.7 \; \rm{cm^{-3}}  \biggl( \frac{6-2\alpha}{5-\alpha}\biggr)^{1/3} \biggl(\frac{R_s}{100 \; \rm{pc}}\biggr)^{-(4+\alpha)/3}\\ 
 \times \; \biggl (\frac{R_0}{100\;\rm{pc}}\biggr)^{\alpha/3} \biggl(\frac{n_{H,0}}{10\; \rm{cm^{-3}}}\biggr)^{1/3} \\ \times \; \biggl(\frac{L_{\rm{kin}}}{5\times 10^{44}\;\rm{erg\;s^{-1}}}\biggr)^{2/3}\biggl(\frac{v_{in}}{0.1c}\biggr)^{-2}.
 \label{eq:n-sw}
\end{multline}
In the second equality in eq. (\ref{eq:n-sw}) we have used the result $R_{sw} \approx R_s (v_s/v_{in})^{1/2}$ appropriate for energy conserving winds.   The shocked wind bubble extends from $R_{sw}$ to the contact discontinuity $R_c$ (Fig. \ref{fig:1}).  The latter is close to the forward shock at $R_s$.   Most of the luminosity of the SW is produced at radii near $R_s$, so we will evaluate the cooling of the SW there.   The density of the shocked wind bubble is roughly constant from $R_{sw}$ to $R_s$ so that $n_{sw}(R_s) \approx n_{sw}(R_{sw})$. 

The existence of a forward and reverse shock separated by a contact discontinuity (as in Fig \ref{fig:1}) requires that the wind have reached the free expansion radius, where the swept up ISM mass is comparable to the total mass ejected in the AGN wind.    This in turn implies that the SAM density is comparable to or larger than the density of the SW given in equation (\ref{eq:n-sw}) (if this is not the case then the AGN wind is over-pressurized relative to the SAM so there is only one shock, not the two-shock structure shown in Fig. \ref{fig:1}).  As a result, free free emission from the SAM always dominates that of the SW outside the free expansion radius.    

Since the SW is in general much hotter (see eq. \ref{eq:T-sw}) and less dense than the SAM, two temperature effects in the post shock region turn out to be important  for the high wind speeds expected (and observed) from AGN (FGQ12).  At the wind shock, protons will be heated to  $T_{p} \approx T_{SW}$ (eq. \ref{eq:T-sw})
but electrons will be heated to a lower temperature (as confirmed by observations of collisionless shocks in the solar wind and supernova remnants; e.g., \citealt{Ghavamian2007}).    Protons and electrons then come into thermal equilibrium via Coulomb collisions, which happens on the timescale \citep{1962pfig.book.....S} \begin{equation}\label{eq:teianalytic}
 t_{ei}=\frac{3m_em_p}{8(2\pi)^{1/2}n_pe^4\rm{ln} \Lambda}\biggl(\frac{kT_e}{m_e}+\frac{kT_p}{m_p}\biggr)^{3/2}
\end{equation}
We use $\ln \Lambda = 40$ in what follows.

Given the low densities in the SW (eq. \ref{eq:n-sw}) and the high photon energy densities close to the AGN, the shocked wind electrons cool primarily by IC emission rather than free-free emission.  

If $t_{ei} > \min(t_{flow},t_{e,IC})$ the electrons and ions do not reach the same temperature (here $t_{e,IC}$ is the time it would take to  the electrons to cool in the absence of Coulomb collisions).   Instead the electrons approach the temperature $T_e^{eq}$ determined by balancing IC cooling and heating from Coulomb collisions with
hot protons (FGQ12):\footnote{If the flow time is very short the electrons may not even reach $T^{eq}_e$ and their temperature would instead be set by the post-shock temperature.   For fiducial parameters, we find that the electrons are within a factor of a few of $T^{eq}_e$ so the equilibrium calculation is sufficiently accurate for our purposes.}
\begin{equation}
 T_e^{eq} \approx 3 \times 10^9 \; \rm{K} \;\times\; \biggl(\frac{v_{in}/v_s}{30}\biggr)^{2/5} \tau_{in}^{2/5}.
 \label{eq:Teq}
\end{equation}
Note that this equation for the electron temperature is only valid so long as $T_e^{eq} \lesssim T_{SW}$.  

Given the SW electron temperature in equation \ref{eq:Teq}, the IC emission from the SAM is
\begin{equation}
\label{eq:L-sw}
 L^{IC}_{SW}\approx L_{\rm{AGN}} \sigma_T N_{SW} \left(1+ \frac{4}{3} <\gamma^2\beta^2>|_{SW} \right)
\end{equation}
where $N_{SW} \approx R_s n_{sw}(R_s)$.
Because $<\gamma^2\beta^2>$ can be somewhat larger than 1 for the SW (up to $\sim 10$ when $R_s \sim 1$ kpc), the transfer of the SW thermal energy to the scattered photons generally dominates over pure Thomson scattering of AGN radiation.   However, so long as $R_s \gtrsim 5$ pc, the SAM IC emission dominates that of the SW.

This is because the density and column density through the SAM are much larger than those through the SW once the forward shock has moved well outside the free expansion radius.

At very small radii, the SW IC emission can be comparable to the total kinetic power of the AGN outflow, but it is generally lower.   This is in contrast to \citet{Bourne2013}, who assumed that $L_{SW}^{IC} \simeq L_{kin}$ even for the two-temperature solutions of FGQ12.   In particular, they argued that because the electron IC cooling time is short, most of the flow energy would be radiated away.   However, most of the shocked wind thermal energy resides in the protons, which only effectively cool on the Coulomb timescale $t_{ei}(T_e^{eq}) \gg t_{flow}$.    As a result, very little of the total thermal energy can be radiated away, consistent with the fact that we find $L_{SW}^{IC} \ll L_{kin}$ under most conditions.

\subsection{Numerical Results}\label{sec:numerics}
In this section we present numerical results for the thermal emission of the shocked ambient medium as a function of radius.  
These are obtained using the equations of \S 3.1-3.3 except that we include first order relativistic corrections in our Compton scattering calculations. Higher order terms are only important for the
SAM within the free expansion radius, where our model is invalid anyway. Results for shock radii less than the free expansion radius are omitted from the plots.
For $n_{H,0}< 10 \, \rm{cm^{-3}}$, $t_{ei}>t_{flow}$ at small radii so that inverse Compton emission is 
somewhat less efficient than the single temperature analytic estimates of \S \ref{sec:IC} would suggest.   We take this into account following the methods of FGQ12. Note that the free-free emission is limited by
the kinetic luminosity of the wind, while the scattered radiation is limited instead by the AGN luminosity and may exceed the kinetic luminosity of the wind.

\begin{figure*}

 \begin{subfigure}{.45\textwidth}

  \includegraphics[scale=.95,keepaspectratio=true,clip=true,trim=90pt 570pt 1pt 1pt]{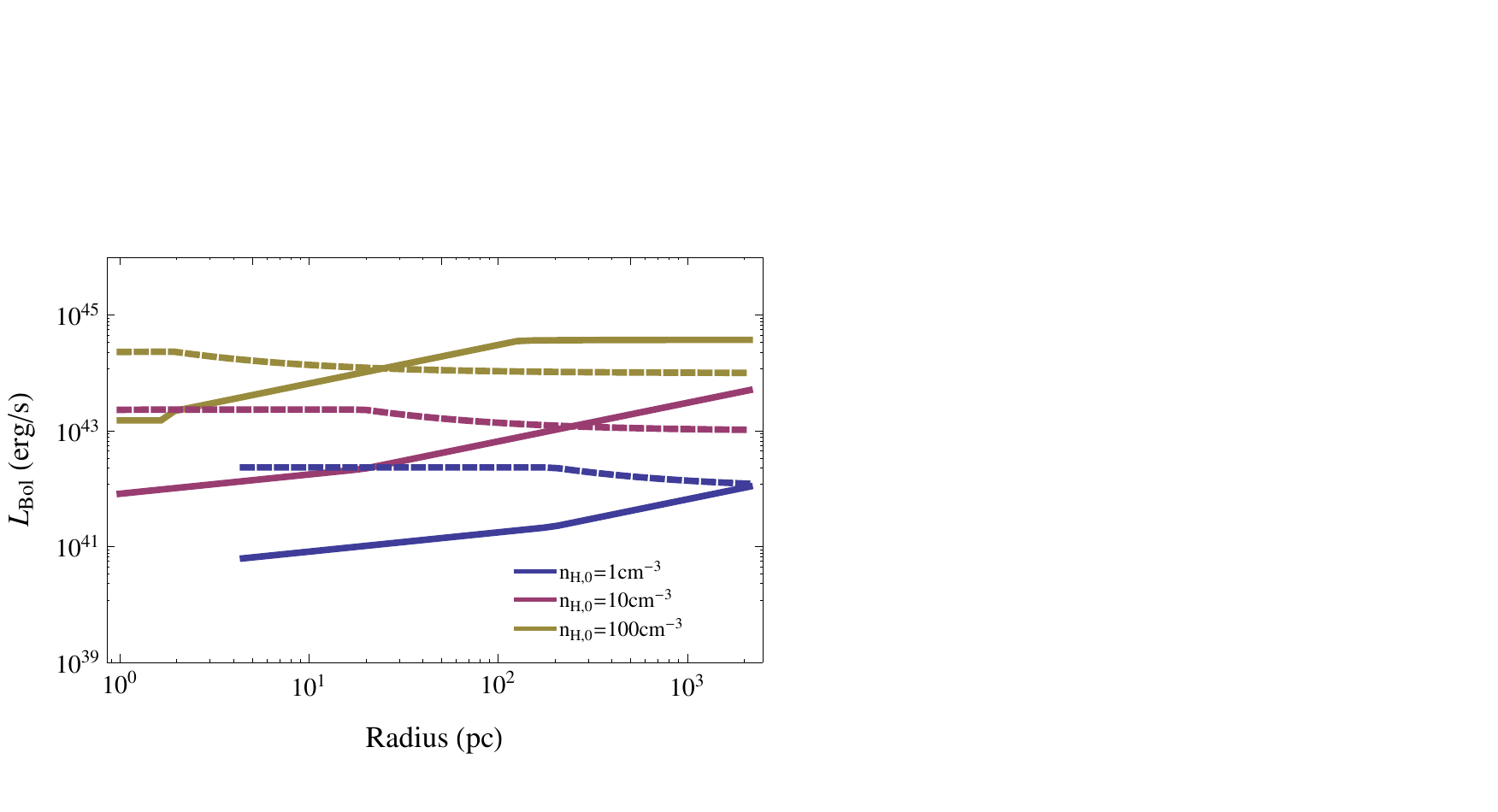}
 
 \end{subfigure}
\begin{subfigure}{.45\textwidth}
  \includegraphics[scale=.95,keepaspectratio=true,clip=true,trim=70pt 570pt 1pt 1pt]{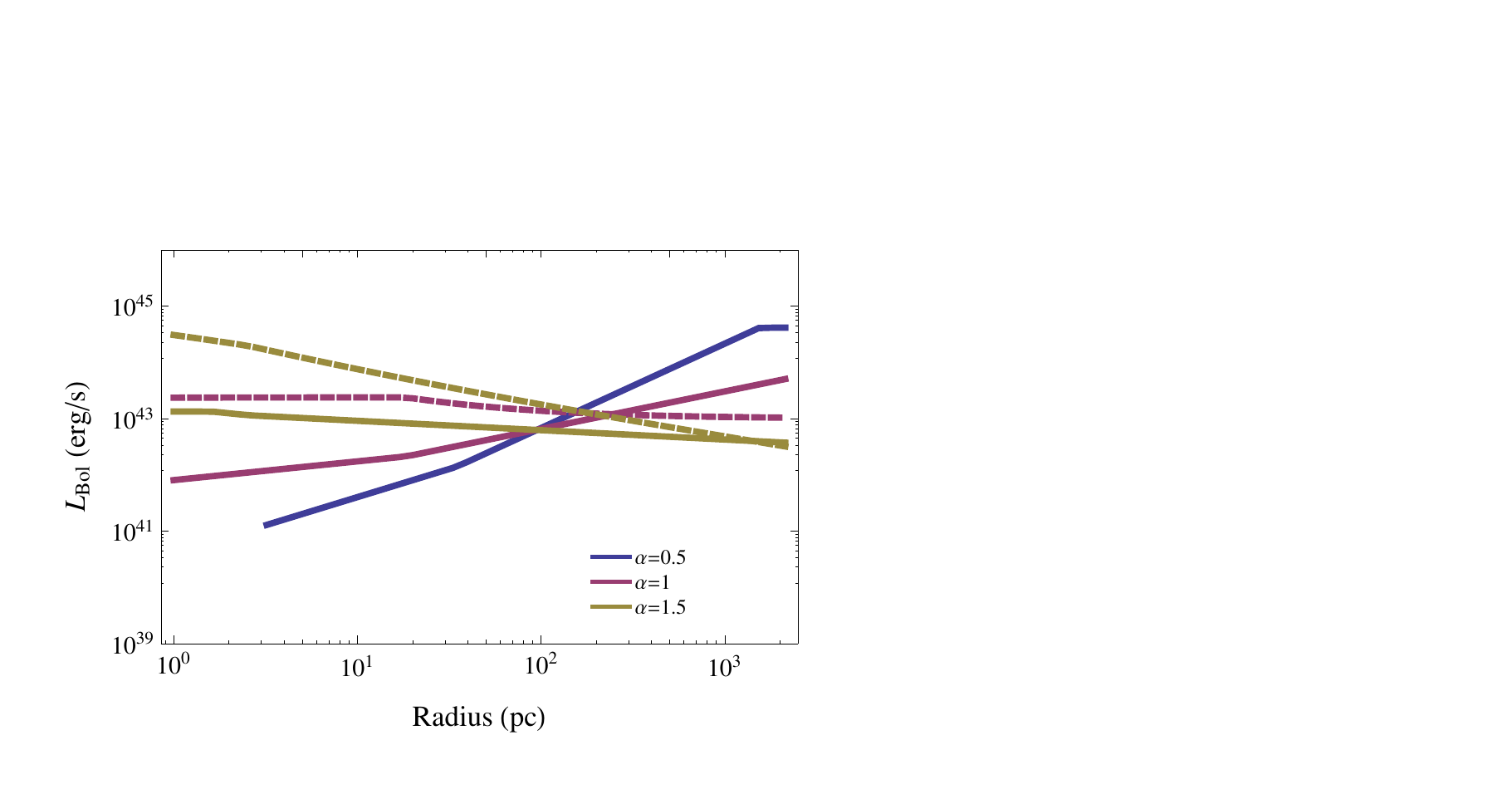}
\end{subfigure}
 \caption{Bolometric luminosity of the ambient interstellar medium shocked by the AGN wind:   free-free emission (solid lines) and IC emission (dashed lines).    {\em Left panel:}  varying $n_{H,0}$ (with $\alpha = 1$), the number density at 100 pc.   {\rm Right panel:}   varying $\alpha$ (with $n_{H,0} = 10$ cm$^{-3}$), the power law index of the ambient medium density profile (see eq. \ref{eq:densityfalloff}). Note that we do not show the $\alpha=0.5$ IC emission because in that case the column density increases without bound as radius increases. The AGN wind is assumed to have $L_{kin}= 5\times 10^{44}$ erg s$^{-1}=0.05L_{\rm{AGN}}$. This is a representative value for the kinetic luminosity of luminous quasars.  }
\label{fig:Lbol}
 \end{figure*}

The left panel of Figure \ref{fig:Lbol} shows the bolometric luminosity vs radius of the forward shock for several values of $n_{H,0}$, fixing $R_0=$100 pc, $\alpha = 1$, and $L_{kin} \approx 5  \times 10^{44}$ erg s$^{-1}$.   The right panel shows the same results for several values of $\alpha$, the power-law slope of the ambient medium density profile (eq. \ref{eq:densityfalloff}).   In all of the plots the solid lines show the free-free emission and the dashed lines show the IC emission.   We only show the IC emission for $\alpha \gtrsim 1$ because otherwise the column density does not converge with increasing radius.   

Figure \ref{fig:Lbol} shows that the SAM cools on a flow timescale at large radii for
 $n_{H,0}\gtrsim 100\;\rm{cm}^{-3}$, but not for lower densities.   In general, IC emission dominates for small radii while free-free dominates at larger radii.   Higher densities (larger $n_{H,0}$) lead to the transition  from inverse Compton to free-free cooling happening at smaller radii since $L_{ff}\propto n_{H,0}^{5/3}$ and $L_{IC}\propto n_{H,0}$. 
 In addition, for $\alpha > 7/5$,  the free-free luminosity is dominated by small radii while for smaller $\alpha$ it is dominated by large radii (see eq. \ref{eq:Lffnocool}).

\begin{figure*}

\begin{subfigure}{.45\textwidth}

    \includegraphics[scale=.95,keepaspectratio=true,clip=true,trim=90pt 570pt 1pt 70pt]{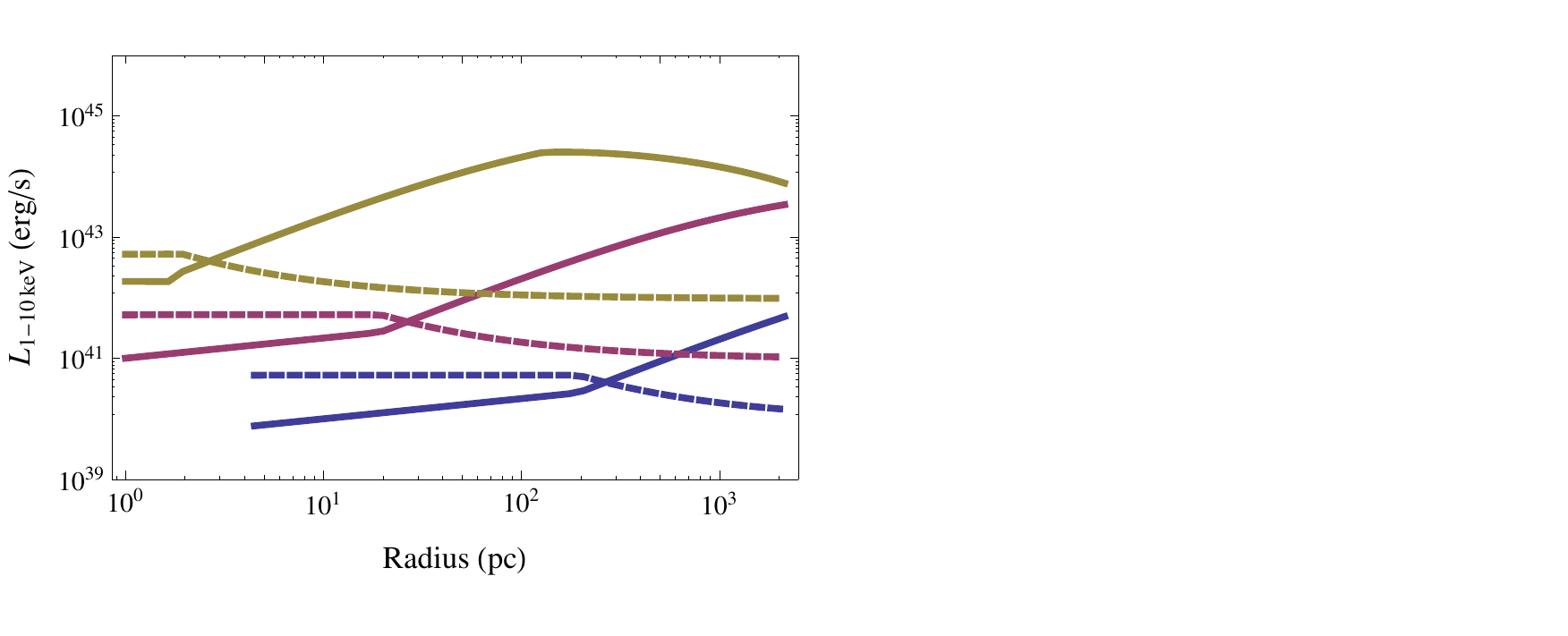}
  
\end{subfigure}
\begin{subfigure}{.45\textwidth}

    \includegraphics[scale=.95,keepaspectratio=true,clip=true,trim=70pt 570pt 1pt 70pt]{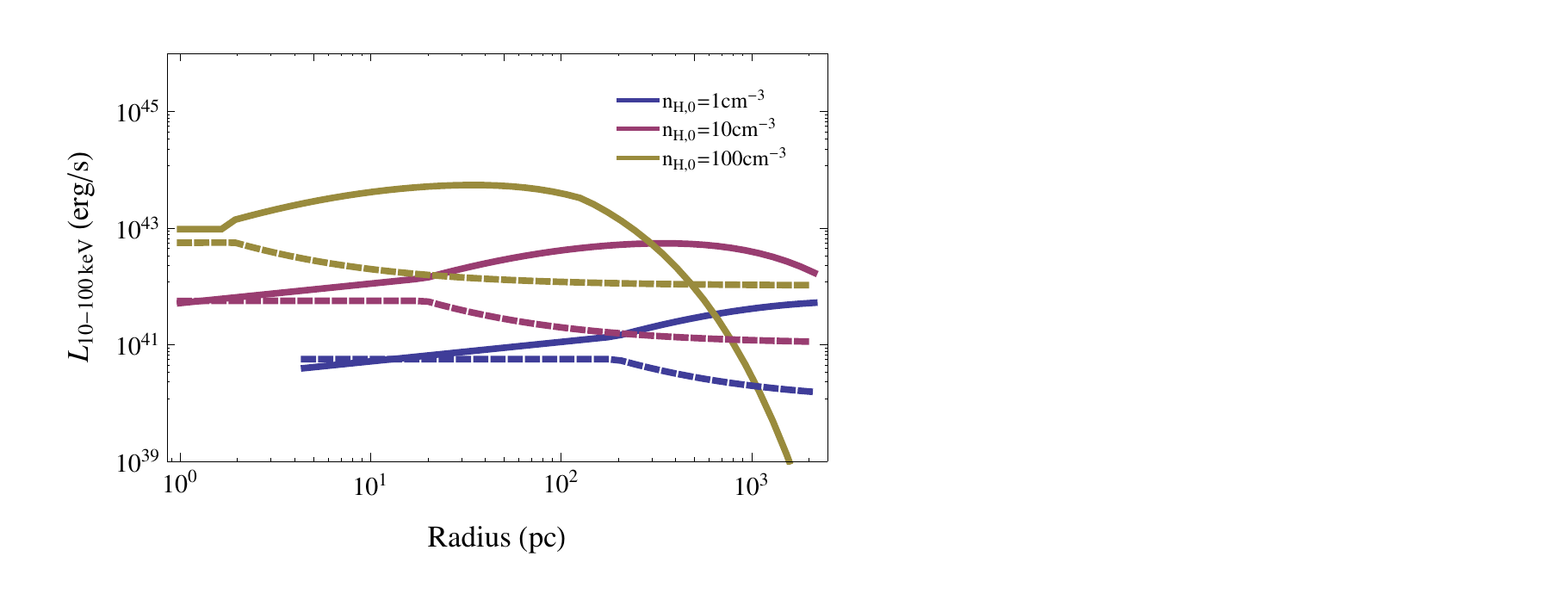}
  
\end{subfigure}

\caption{X-ray luminosities produced by the shocked ambient medium in 1-10 keV ({\em Left}) and 10-100 keV ({\em Right}) bands, for varying $n_{H,0}$; $\alpha = 1$ and $L_{kin} \approx 5\times 10^{44}$ erg s$^{-1}$ are fixed.    Solid lines show free-free emission while dashed lines show IC emission.  For $n_{H,0} \gtrsim 1 \cc$, free-free emission  dominates the X-ray emission.   In all cases the $1-10$ keV emission is the best probe of the bolometric luminosity shown in Figure \ref{fig:Lbol}.  Most of the X-ray emission is produced at large radii, except for the highest density case with $n_{H,0}\sim 100 \; \rm{cm^{-3}}$.  In this case, the shock decelerates rapidly and  the emission shifts to longer wavelengths. }

\label{fig:xraynH}
\end{figure*}

\begin{figure*}
\begin{subfigure}{.45\textwidth}

    \includegraphics[scale=.95,keepaspectratio=true,clip=true,trim=90pt 570pt 1pt 70pt]{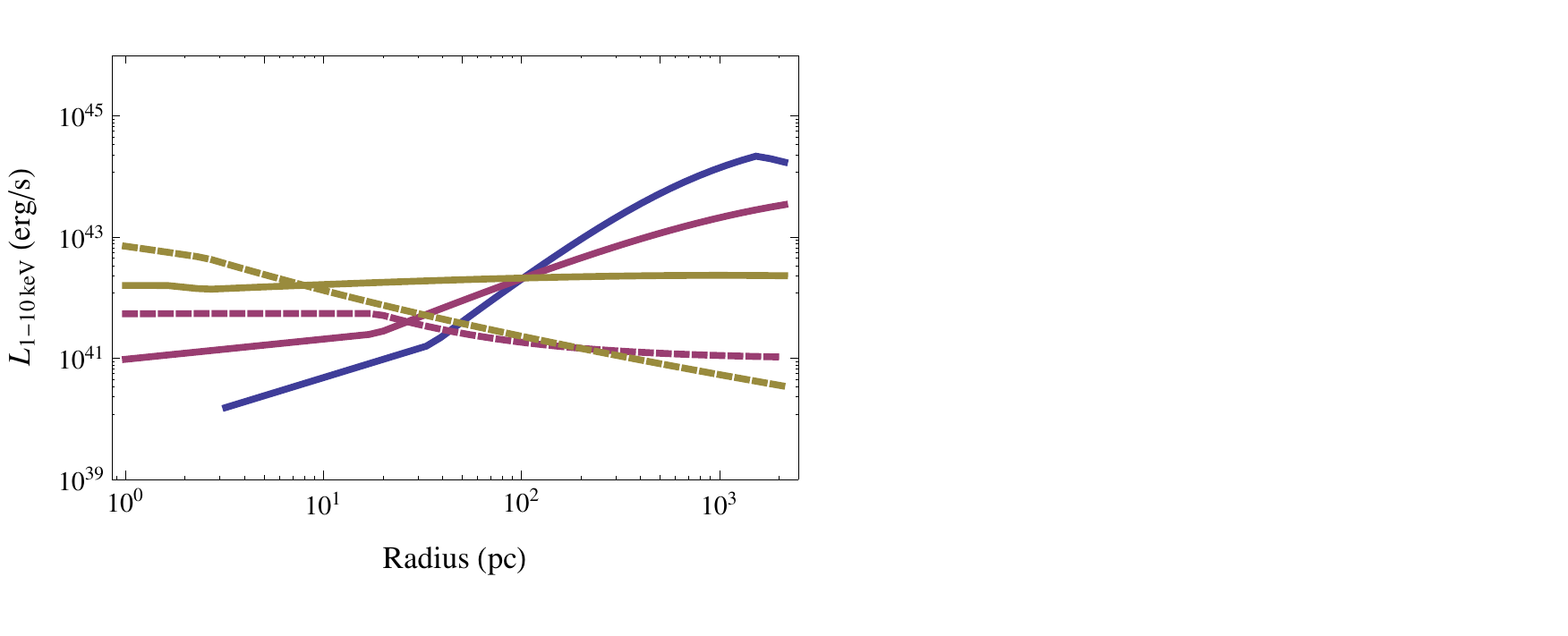}
\end{subfigure}
\begin{subfigure}{.45\textwidth}

    \includegraphics[scale=.95,keepaspectratio=true,clip=true,trim=70pt 570pt 1pt 70pt]{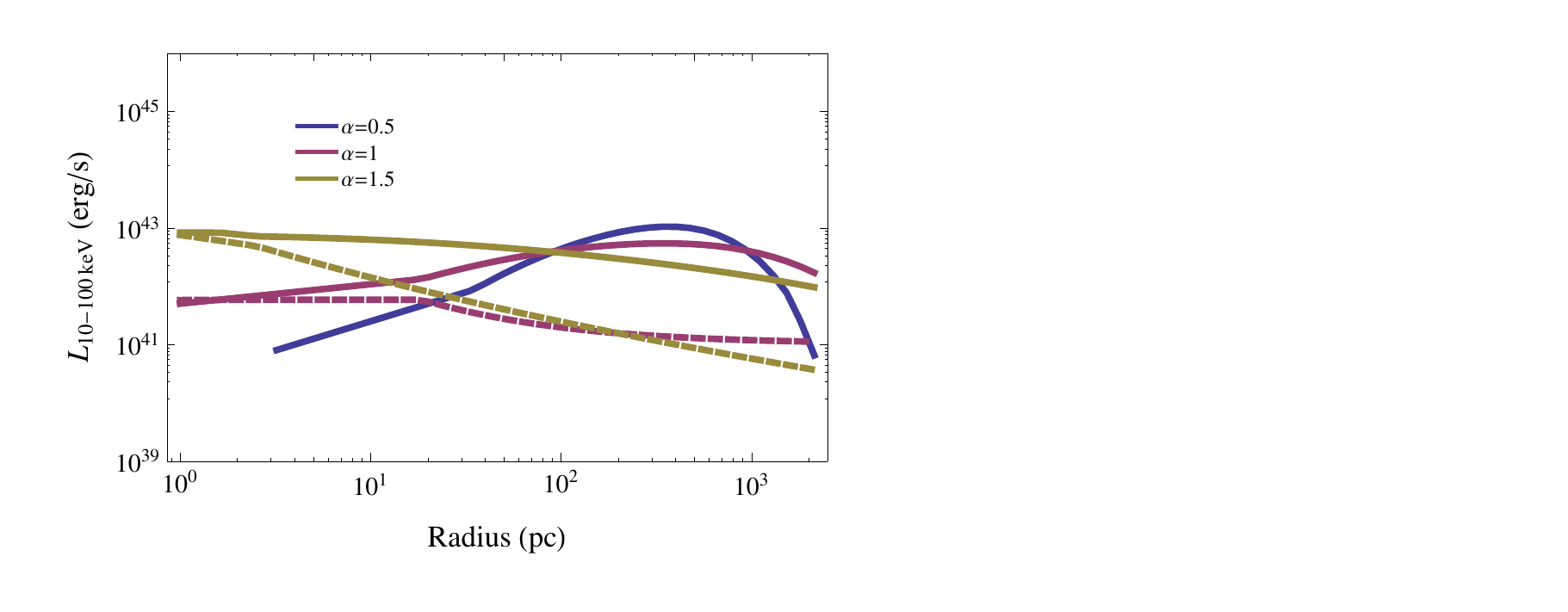}
\end{subfigure}
\caption{X-ray luminosities produced by the shocked ambient medium in 1-10 keV ({\em Left}) and 10-100 keV ({\em Right}) bands for different values of $\alpha$, the power law index of the ambient medium density profile (see eq. \ref{eq:densityfalloff}); the density at 100 pc is $n_{H,0} = 10$ cm$^{-3}$ in all cases and the AGN wind has $L_{kin} \approx 5\times 10^{44}$ erg s$^{-1}$.  Solid lines show free-free emission while dashed lines show IC emission.  We do not plot the  IC emission for $\alpha=0.5$  because in that case the column density increases without bound as radius increases.}

\label{fig:xrayAlpha}
\end{figure*}

Figures  \ref{fig:xraynH} and \ref{fig:xrayAlpha} show the radiation from the SAM in the 1-10 keV and 10-100 keV bands varying $n_{H,0}$ and $\alpha$, respectively.  
Free-free spectra are obtained using the non-relativistic free-free emissivity from \citet{1979rpa..book.....R}. 
For the IC spectrum, use the the input AGN spectrum in equation (\ref{eq:imputspectrum}) and
 assume pure Thomson scattering (because $kT_{SAM} \ll m_e c^2$ is appropriate for most of the SAM; see \S \ref{sec:IC}).

To interpret these results it is useful to note that  equations \ref{eq:SAMtemp} \& \ref{eq:vs}  imply that for $n_{H,0} = 10$ cm$^{-3}$ and $\alpha = 1$, the temperature of the SAM decreases from $\sim 3 \times 10^8$ K at $R_s = 100$ pc to $\sim 6 \times 10^7$ K at $R_s = 1$ kpc.   For other $n_{H,0}$, $T_{SAM} \propto n_{H,0}^{-2/3}$.   Thus much of the free-free emission at large radii comes out in the $\sim 1-10$ keV x-rays (see also eq. \ref{eq:LX}).   This is not true at smaller radii because the post shock temperatures are higher, but inverse Compton emission dominates at smaller radii regardless.

The numerical results in Figures  \ref{fig:xraynH} and \ref{fig:xrayAlpha} confirm that a significant fraction of the bolometric luminosity is radiated in the $1-10$ keV band, particularly at larger radii $\sim 0.1-1$ kpc where the shock has decelerated somewhat.   Higher ambient densities result in faster deceleration, suppressing the hard x-ray emission from large radii in the case of $n_{H,0} = 100$ cm$^{-3}$ and $\alpha = 1$.  

\subsection{Comparison to X-ray Emission Produced by Star Formation}\label{sec:starform}

Observations of star forming galaxies show that there is a reasonably tight correlation between the spatially extended X-ray emission and the star formation rate \citep{Mineo2014}: 

$L_{\rm X(0.5-8keV)} \approx 4 \times 10^{39} (\dot M_*/1 \ M_\odot \, {\rm yr^{-1}}) \, {\rm erg \, s^{-1}}$.
Most of this emission is produced by high mass X-ray binaries.   The contribution from diffuse hot gas is $\sim 10\%$ in a few nearby cases where the different components contributing to the X-ray emission can be robustly identified (e.g., M82; \citealt{Strickland2009}).   

The rate of star formation  in the host galaxies of luminous quasars is not fully understood (compare, e.g., \citealt{Ho2005,Condon2013,2014MNRAS.442..784Z}).   However, even if $\dot M_* \sim 300 \mpy$ (which corresponds to $L_* \sim L_{\rm AGN}$ for our fiducial $L_{\rm AGN} \sim 10^{46} \erg$), the X-ray luminosity associated with star formation is $L_X \sim 10^{42} \erg$.   This is  less than the $1-10$ keV X-ray luminosity we predict for the shocked ambient medium at radii $\sim 0.1-1$ kpc for $n_{H,0}\gtrsim 10 \rm{cm^{-3}}$. (Fig. \ref{fig:xraynH} \& \ref{fig:xrayAlpha}).   In particular, models in which the AGN wind decelerates significantly to speeds of $\sim 1000 \kms$ via interaction with ambient gas produce significantly brighter X-ray emission.  This strongly suggests that AGN powered galactic winds will be associated with spatially extended diffuse X-ray emission significantly in excess of that typical for star forming galaxies.   Moreover, it is quite possible that the X-ray emission predicted here {\em underestimates} the total X-ray emission from the galactic wind.  Studies of nearby star formation powered galactic winds suggest that much of the X-ray emission is produced by the interface regions between a hot flow and cool entrained gas (e.g., \citealt{Veilleux2005}).   The hot flow itself, which is essentially what we are modeling here, is much more difficult to directly detect.

\section{Non-thermal Emission}\label{sec:nonthermal}
We now analyze emission from non-thermal electrons accelerated by AGN wind shocks.   Both the wind shock and the forward shock in the ISM are likely sites of particle acceleration and associated non-thermal emission.
    
We assume that shock accelerated electrons are produced with a power-law distribution
 $\dot{n}(\gamma)=A\gamma^{-p}$ where A is a normalization constant. 
 A power law index  $p \approx 2$ or slightly above 2 is consistent with shock acceleration theory and observations of supernova remnants \citep{1987PhR...154....1B}; we take $p = 2$ in what follows.    The normalization constant $A$ is determined by the total energy supplied to relativistic electrons at the shock, which we take to be a fraction $\xi$ of the total kinetic energy flux in the AGN wind:
\begin{equation}
\dot{E}_{e}= \int_1^{\gamma_{max}} \rm{d}\gamma A\gamma^{-2}\gamma m_ec^2 = 10^{-2}\xi_{-2}L_{kin}.
\end{equation}
We have scaled to $\xi = 10^{-2} \xi_{-2}$, i.e.,  1\% of the shock energy going into relativistic electrons. This is required by observations of the total non-thermal synchrotron emission from galaxies, which measures the integrated effect of electron acceleration at supernova shocks \citep{2006ApJ...645..186T}.   In particular, there is no obvious difference in the properties of AGN-driven shocks relative to supernova shocks that would  change the average electron acceleration efficiency.

The ratio of non-thermal synchrotron to inverse Compton emission is given by
\begin{multline}\label{eq:ratio}
 \frac{P_{synch}}{P_{IC}}= \frac{U_B}{U_{ph}} \\ \approx  1 \, \biggl(\frac{B}{3 \; \rm{mG}}\biggr)^2 \biggl(\frac{R_s}{100\;\rm{pc}}\biggr)^2\biggl(\frac{10^{46}\; \rm{erg\; s^{-1}}}{L_{\rm{AGN}}}\biggr).
\end{multline}
In equation (\ref{eq:ratio}), we have used a fiducial magnetic field strength of 3 mG, assuming shock compression of an ambient $\sim$ mG magnetic field; the latter is suggested by 
observations of high star formation rate galaxies \citep{1991ApJ...378...65C,2014ApJ...780..182M}.  
Equation (\ref{eq:ratio}) shows that considering just the shock compressed ISM magnetic field, synchrotron emission will dominate inverse Compton emission for radii $\gtrsim 0.1 \;\rm{kpc}$. 
It is possible that synchrotron emission dominates at even smaller radii if there is efficient magnetic amplification in the vicinity of the shock.   

\subsection{Synchrotron Emission}
The timescale for synchrotron cooling by an electron with Lorentz factor $\gamma$ is
\begin{equation}
t_{synch} \approx 3 \times 10^{6}\; \rm{yr}\times\biggl(\frac{3\,\rm{mG}}{B}\biggr)^2\gamma^{-1}. 
\end{equation}
Comparing this to $t_{flow}$ allows us to find the critical $\gamma$ above which electrons cool.

We can equivalently define a synchrotron cooling frequency above which electrons will radiate all of the energy provided by shock acceleration:
\begin{equation}
 \nu_{\rm cool} \approx 10 \; \rm{MHz} \; \biggl(\frac{v_s}{1000 \; \rm{km\;s^{-1}}}\biggr)^2\biggl(\frac{3\; \rm{mG}}{B}\biggr)^3\biggl(\frac{100 \, pc}{R_s}\biggr)^2.
\end{equation}
For $\nu \gtrsim \nu_{\rm cool}$, the predicted emission is
\begin{multline}
\nu L_{\nu} \approx \frac{10^{-2} \, \xi_{-2} \, L_{kin}}{2\; \rm{ln}(\gamma_{max})}   \ \  \ \ ({\rm \small synchrotron: \ radio \ to \ X-ray}) \\ \approx 10^{-5} \, \xi_{-2} \, L_{\rm AGN} \, \left(\frac{L_{kin}}{0.05 \, L_{\rm AGN}}\right) \ \ \ \ (\nu \gtrsim \nu_{\rm cool})
\label{eq:Lsynch}
\end{multline}
The flat spectrum in $\nu L_\nu$ for $\nu \gtrsim \nu_{\rm cool}$ is a consequence of our choice of $\dot{n}(\gamma)\propto \gamma^{-2}$, which corresponds to equal energy per logarithmic interval in electron energy.
Physically, as long as the magnetic field is
sufficiently strong  that synchrotron dominates over IC and electrons with a given energy cool on the flow time, the synchrotron emission is {\em independent} of the magnetic field strength and the input electron spectrum with equal energy per logarithmic interval of electron energy translates into a flat radio spectrum in $\nu L_\nu$.
For $\nu \lesssim \nu_{\rm cool},  \nu L_\nu \propto \nu^{1/2}$.

Equation (\ref{eq:Lsynch}) is much less than the radio emission from radio loud quasars, which is  $\sim 10^{-4} L_{\rm AGN}$  \citep{2014MNRAS.438.2253S}.   However, our predicted radio flux for AGN winds is of order the emission from radio quiet quasars.    This suggests that observations of radio quiet AGN provide one of the best quantitative probes of the energy content of AGN outflows.    Two clear observational ways of distinguishing between jet-dominated and wind-dominated radio emission are (1) the spatial extent of the emission and (2) the spectral indices, which should be (by the same logic used above to explain equation \ref{eq:Lsynch}) $F_\nu \sim \nu^{-b}$, with $b \sim 1$ for wind dominated emission (slightly steeper if $p > 2$).   A more challenging task is distinguishing between AGN and star formation powered radio emission in radio-quiet quasars.  The reason is that equation \ref{eq:Lsynch} is not that dissimilar from the standard FIR-radio correlation of star forming galaxies, which is $\nu L_\nu (\nu = 1.4 {\rm GHz}) \approx 2 \times 10^{-6} L_{FIR}$ \citep{Yun2001}.   Indeed it is well known that luminous radio quiet Seyferts and radio quiet quasars lie roughly on the FIR-radio correlation (e.g., \citealt{Roy1998,Condon2013}).   

Testing equation (\ref{eq:Lsynch}) ideally requires correlating the radio emission from radio quiet quasars with other wavelengths that best track the bolometric AGN luminosity.   Far UV, mid IR, and X-ray are particularly promising possibilities because of the modest scatter in emission at these wavelengths for AGN of similar bolometric luminosity \citep{2014MNRAS.438.2253S}.   By contrast, because the FIR emission can include a significant contribution from star formation, we expect that if shocks driven by AGN winds in radio quiet quasars indeed dominate their radio emission, the radio emission may correlate less well (i.e., with more scatter) with the FIR emission than with these more direct tracers of the AGN's bolometric luminosity.

The synchrotron emission predicted by equation (\ref{eq:Lsynch}) will extend into the high energy bands if electrons with very high Lorentz factors are accelerated.    The maximum Lorentz factor of the electrons can be roughly estimated by setting the synchrotron cooling time comparable to the timescale for diffusive shock acceleration, which is $t_{acc} \sim \Omega_c^{-1} c^2/v_s^2$ \citep{1987PhR...154....1B} where $\Omega_c$ is the relativistic cyclotron frequency.   This yields a maximum photon energy for synchrotron emission of
\begin{equation} \label{eq:Emax}
E^{synch}_{\rm max} \sim 10 \, {\rm keV} \;\times\; \left(\frac{v_s}{1000 \ {\rm km \, s^{-1}}}\right)^2
\end{equation}
where we have continued to assume that synchrotron is the dominant coolant for relativistic electrons.   For a fiducial AGN luminosity of $L_{\rm AGN} = 10^{46}$ erg s$^{-1}$, equation (\ref{eq:Lsynch}) predicts a non-thermal synchrotron $\sim 1-10$ keV X-ray emission of $\sim 10^{41}$ erg s$^{-1}$.   Except for the lowest densities ($n_{H,0} \lesssim 1$ cm$^{-3}$), this is less than the thermal free-free emission predicted by equation (\ref{eq:Lffnocool}) and Figures \ref{fig:xraynH} and \ref{fig:xrayAlpha}.  There will also be inverse Compton $\gamma$-ray emission from the nonthermal electrons but at large radii where $U_B > U_{ph}$, this is less than the synchrotron emission predicted by equation (\ref{eq:Lsynch}) by a factor of $U_B/U_{ph}$.  The inverse Compton emission will, however, extend to higher energies $\sim$ TeV.       

\subsection{Inverse Compton Emission}
At small radii where $U_{ph} \gtrsim U_B$ (eq. \ref{eq:ratio}), inverse Compton cooling will be the dominant energy loss mechanism for accelerated electrons.    The non-thermal inverse Compton emission is then
\begin{align}
\nu L_{\nu} \approx  \frac{0.01 \, \xi_{-2} \, L_{kin}}{2\, \rm{ln}(\gamma_{max})}   \ \  \ \ ({\rm \small IC:  \ E^{IC}_{min} \ to \ E^{IC}_{max} })  \nonumber \\ \approx 10^{-5} \, \xi_{-2} \, L_{\rm AGN} \, \left(\frac{L_{kin}}{0.05 \, L_{\rm AGN}}\right) 
\label{eq:LIC}
\end{align}
We can analytically estimate $E^{IC}_{min}$ by determining the Lorentz factor $\gamma_{min}$ above which  the IC cooling timescale is less than the flow timescale and using the fact that the final photon energy is $\sim \gamma_{min}^2$ of the input photon energy $E_{min}^{AGN}$:
\begin{multline}
 E^{IC}_{min}\sim 1 \; \rm{keV} \; \biggl(\frac{L_{\rm{AGN}}}{10^{46}\; \rm{erg\;s^{-1}}}\biggr)^{-1} \, \biggl(\frac{R_s}{100 \; \rm{pc}}\biggr) \\ \times \biggl(\frac{v_s}{1000 \; \kms}\biggr) \, \biggl(\frac{E_{\rm min}^{\rm AGN}}{1 \, {\rm eV}}\biggr)
\end{multline}
The maximum energy of the non-thermal IC photons $E^{IC}_{max}$ can be estimated by using the fact that   Klein-Nishina corrections suppress the cross section for photon energies $\gtrsim m_e c^2$ in the electron rest frame.   This yields
\begin{equation}
 E^{IC}_{max} \sim 0.3 \; \rm{TeV} \, \biggl(\frac{E_{min}^{AGN}}{1 \, {\rm eV}}\biggr)^{-1}
\end{equation}
Because AGN emit at a wide range of energies, including at $\lesssim 1$ eV, it is likely that the non-thermal IC emission extends from the soft X-rays well into the TeV $\gamma$-rays.

The non-thermal inverse Compton emission predicted by equation (\ref{eq:LIC}) is $\sim 10^{41} \erg$ for $L_{\rm AGN} \sim 10^{46} \erg$.   This is typically less than the thermal hard X-ray contribution  in Figures \ref{fig:xraynH} and \ref{fig:xrayAlpha} except for the lowest ambient densities $n_{H,0} \sim 1 \, {\rm cm^{-3}}$.  The nonthermal IC hard X-ray emission will, however, extend to significantly higher energies.

\subsection{$\gamma$-rays From Pion Decay}

Finally, we note that under certain circumstances, $\gamma$-ray emission from the decay of neutral pions created by relativistic protons accelerated at the AGN forward shock may produce a significant $\gamma$-ray luminosity at $\gtrsim 1$ GeV.   This can be energetically more important than the leptonic $\gamma$-ray emission from inverse Compton emission.   The maximum luminosity from pion decay is set by the energy supplied to relativistic protons at the forward shock, $\dot E_p \approx \eta L_{kin}$, where $\eta \sim 10 \, \xi \sim 0.1 \eta_{-1}$, i.e., we assume that the protons receive
$\sim 10$ times the energy supplied to relativistic electrons (see \citealt{2005AN....326..414B}).    In the limit that  the relativistic protons lose the majority of their energy to pion production before leaving the galaxy the resulting $\gtrsim$ GeV $\gamma$-ray luminosity is $\sim 10$ times larger than equation (\ref{eq:Lsynch}), i.e., 
\begin{equation}
\nu L_\nu \approx 2 \times 10^{-4} \, \eta_{-1} \, L_{\rm AGN} \, \left(\frac{L_{kin}}{0.05 \, L_{\rm AGN}}\right) \ \ \ \ (E \gtrsim 1 {\rm GeV}).
\label{eq:pion}
\end{equation}
Equation (\ref{eq:pion}) predicts a $\gamma$-ray luminosity of $\sim 10^{42}$ erg s$^{-1}$ for a typical 
$L_{\rm AGN} = 10^{46}$ erg s$^{-1}$.  We reiterate that this is the {\em maximum} $\gamma$-ray emission from pion decay.   Although it is quite uncertain whether this maximum luminosity is realized, the nearby starburst galaxies M82 and NGC 253 do roughly conform (to a factor of $\sim 3$) to the analogous prediction for the maximum $\gamma$-ray emission associated with  relativistic protons accelerated at supernova shocks interacting with the dense ISM prior to escaping the galaxy \citep{Lacki2011}.

\section{Discussion}
\label{sec:discussion}

We have examined the thermal and nonthermal emission produced by galactic winds powered by active galactic nuclei (AGN).    Most of the emission is produced by the shocked ambient medium (SAM) created as a forward shock is driven into the surrounding interstellar medium (Fig. \ref{fig:1}).  
In particular, thermal emission from the SAM significantly exceeds that of the shocked AGN wind itself (\S \ref{sec:SW}).   It is, however, cooling of the shocked AGN wind that determines whether AGN-powered galactic winds conserve energy or momentum (FGQ12).   Rapid cooling of the SAM -- and the associated high luminosity of an AGN-powered galactic wind -- thus does not necessarily imply that the outflow conserves momentum rather than energy (in contrast to some claims; e.g.,  \citealt{Pounds2011}).

Our models  assume that AGN-powered galactic winds have kinetic luminosities $\sim 0.01-0.1 \, L_{\rm AGN}$ and momentum fluxes of $\sim 1-10 \, L_{\rm AGN}/c$. These values are motivated by existing observations  (e.g., \citealt{2012MNRAS.420.1347F,Cicone2014}) and by models of the $M_{BH}-\sigma$ relation.   
We further assume that the AGN wind interacts with ISM gas that is sufficiently dense to decelerate the AGN outflow on scales of $\sim 1$ kpc.  If it does not, then the AGN wind simply `vents.'  Thus our models are intended to make predictions that can  test the hypothesis that AGN feedback has a significant effect on gas in the ISM of the AGN's host galaxy.

Our key results include the following:

\noindent{\bf 1.}   The SAM at radii $\sim 0.1-3$ kpc should produce a spatially extended 1-10 keV X-ray luminosity from free-free emission of $\sim 10^{41}-10^{44} \erg$ with the exact value depending on the shock radius and the ambient ISM density (Fig. \ref{fig:xraynH} \& \ref{fig:xrayAlpha}).  This significantly exceeds the spatially extended X-ray emission of star forming galaxies (\S \ref{sec:starform}), providing a possible way to observationally distinguish between AGN and star formation powered galactic winds.   The thermal free-free emission from the SAM depends primarily on the density of the ambient medium shocked by the AGN wind, and only weakly on the parameters of the AGN wind itself (eq. \ref{eq:LffNoCool} \& \ref{eq:LX}).
The thermal emission from the SAM is nonetheless an excellent probe of whether the AGN wind interacts significantly with the ISM of the host galaxy, which is a pre-requisite for efficient AGN feedback. 

\noindent{\bf 2.}   Shock accelerated relativistic electrons produce non-thermal synchrotron emission from the radio to the soft X-rays with $\nu L_\nu \sim 10^{-5} L_{\rm AGN}$ (eq. \ref{eq:Lsynch}) at radii $\gtrsim 100$ pc.    At smaller radii, inverse Compton cooling by the AGN's radiation likely suppresses the synchrotron emission relative to this prediction, though the precise degree of this suppression depends on the uncertain amplification of ISM magnetic fields in the vicinity of the shock.    
At small radii $\lesssim 100$ pc where inverse Compton emission dominates, the non-thermal inverse Compton  luminosity is $\sim 10^{41} \erg  \, (L_{\rm AGN}/10^{46} \erg)$ (eq. \ref{eq:LIC}) from soft X-rays to TeV $\gamma$-rays.    The $\gtrsim$ GeV $\gamma$-ray emission can be up to a factor of $\sim 10$ brighter $\sim 10^{42} \erg \, (L_{\rm AGN}/10^{46} \erg) $ due to emission from the decay of neutral pions created as shock accelerated protons interact with the ambient ISM.

We now discuss our results in the context of current observational constraints.  \cite{2014ApJ...788...54G} observed a Type II quasar with  $\sim 20 \;\rm{kpc}$ bubble of diffuse X-ray emission having a temperature of $3\times 10^6 \;\rm{K}$ and $L_{0.3-8\;keV} \sim 10^{41} \; \rm{erg \;s^{-1}}$.  This is roughly the size expected for a bubble produced during a quasar's $\sim 10^7$ yr lifetime (eq. \ref{eq:Rs}).  The observed temperature, interpreted as the temperature of gas at the forward shock ($T_{SAM}$; eq. \ref{eq:SAMtemp}), suggests a shock velocity of $\sim 400 \; \rm{km \; s^{-1}}$.   A more likely interpretation is that the soft X-ray emission is produced at the interface between a faster underlying quasar-powered galactic outflow and denser cool clouds  (see \S \ref{sec:starform} and the discussion below).

The non-thermal radio emission we predict from AGN-powered galactic winds (eq. \ref{eq:Lsynch}) is much less than that observed in radio-loud quasars.   It is, however, comparable to the radio emission observed in radio-quiet quasars.   Radio quiet quasars also lie roughly on the high luminosity extension of the FIR-radio correlation of star forming galaxies \citep{Roy1998}.    This either implies that both the FIR and radio emission in radio-quiet quasars are dominated by star formation (e.g., \citealt{Condon2013}) or that  radio quiet quasars and star formation coincidentally produce similar ratios of radio and FIR emission  (e.g., \citealt{2014MNRAS.442..784Z}).   Our results  lend some support to the latter hypothesis (see \S \ref{sec:nonthermal}).
Specifically, we have shown that if  quasars drive outflows with energetics suggested by both independent observations and theoretical models of black hole self-regulation, then they should produce radio emission with $\nu L_\nu \sim 10^{-5} L_{\rm AGN}$ (at least when the outflows reach radii $\gtrsim 100$ pc where synchrotron dominates over inverse Compton losses; see eq. \ref{eq:Lsynch} and related discussion).  
This prediction is not that different from the observed FIR-radio correlation, which is $\nu L_\nu(1.4 {\rm \, GHz}) \approx 2 \times 10^{-6} L_{\rm FIR}$.   Ultimately, this is due to the fact that the kinetic power in AGN outflows suggested observationally and theoretically ($\sim 0.01-0.1 \, L_{\rm AGN}$) is rather similar to the energy supplied to the ISM by supernova shocks ($\approx 10^{-2} L_\star$).  Thus the energy budget available for accelerating relativistic electrons is plausibly similar in both the star formation and AGN-wind cases.  

There are several ways to test the AGN-wind origin of the radio emission in radio quiet quasars.  One is to specifically test whether the radio emission  correlates better with observational indicators of star formation or observational indicators of AGN activity (e.g., hard X-rays, narrow line emission).   \citet{2014MNRAS.442..784Z} carried out such a test and concluded that the correlation between radio-quiet quasar radio emission and 5007$\;$\AA$\;$   O[III] line velocity favored an AGN  origin. \citet{2013MNRAS.433..622M} also showed that there is
a correlation  between the 1.4 GHz radio emission and the 5007$\;$\AA$\;$  O[III] line width and several other  AGN luminosity measures.     A second test of the AGN-wind origin of the radio emission in radio quiet quasars is to study the most luminous quasars. These have $L_{\rm AGN}$ larger than the star formation luminosity produced by any known population of star forming galaxies.   
If the radio emission traces AGN activity then this luminous quasar sub-sample might well still have $\nu L_\nu \sim 10^{-5} \, L_{\rm AGN}$ while if it traces star formation this sub-sample should have systematically lower radio luminosities.   One subtlety in carrying out these tests is that the radio emission from AGN winds is a measure of the integrated energy supplied by the quasar to a galactic wind on timescales comparable to the synchrotron cooling timescale, while many other AGN indicators (e.g., hard X-rays) trace the instantaneous accretion rate.   Large variability in the latter could complicate these comparisons.

Firmly identifying the origin of the radio emission in radio quiet quasars has the potential to   provide one of the best probes of the impact of AGN-powered galactic winds on their surrounding  galaxy.   Specifically, our results show that a careful compilation of $\nu L_\nu/L_{\rm AGN}$ for a population of radio quiet quasars will directly constrain the average kinetic power in quasar outflows (i.e., $L_{kin}$ in eq. \ref{eq:Lsynch}).

The biggest simplification in our calculations is that we model the AGN outflow and the surrounding medium as homogeneous and spherically symmetric.   It is possible, of course, that the AGN wind simply vents out of the polar regions of the galaxy and has little effect on the interstellar gas (e.g., \citealt{2013MNRAS.434..606G,2014MNRAS.440.2625Z}).   Our models are specifically intended to test the alternate possibility in which the AGN wind {\em does} interact significantly with the ISM.  This is appropriate if the gas disk in the host galaxy subtends a significant solid angle viewed from the AGN, e.g., due to feedback from star formation or dynamical processes such as disk warps.  Even in this case, however, the ISM through which an AGN wind propagates will in reality be inhomogeneous, with most of the volume   occupied by gas at densities well below the mean, while most of the mass is contained in a relatively small volume (``clouds").    Our calculations describe the emission produced by interaction between the AGN wind and gas filling most of the volume.  
The bulk of the emission produced by interaction with dense clouds is likely to be  more difficult to detect.   In particular, if the AGN wind encounters clouds with densities $n_{cl}/n_H \gg 1$ (where $n_H$ is the density of the volume filling gas), the clouds will be shocked to a temperature $\sim T_{\rm SAM} (n_H/n_{cl})$.   As argued in \S \ref{sec:setup} the mean gas densities in the central $\sim 0.1$ kpc of gas-rich galaxies are a factor of $\sim 10-100$ larger than the typical gas densities the AGN winds can sweep up given speeds of $\sim 1000 \kms$ at radii $\sim 0.1-1$ kpc.      This suggests that AGN winds will inevitably encounter clouds with densities $\gtrsim 100$ times larger than the typical density of the volume filling medium.  The temperature such clouds will be shocked to is $\sim 10^{5-6}$ K and most of the radiation will thus come out in UV lines, which will be reprocessed into the infrared by  dust.  The bulk of the emission from shocked clouds will be very difficult to directly observe.   However, the high velocity molecular gas observed in Mrk 231 and nearby ULIRGs (e.g., \citealt{2010A&A...518L.155F}) is presumably a signature of this interaction.   Moreover, much of the spatially extended X-ray emission may be produced at the interface between the AGN wind and cool clouds.

In the context of the simple estimates in this paper, the primary uncertainty to highlight is that
the density that comes into the emission predictions for the bulk of the {\em volume} of the SAM can differ from the density that determines the dynamics of the wind as it decelerates.   This will be the case  if most of the mass that contributes to decelerating the AGN wind is initially in overdense clouds that efficiently mix into the AGN wind.   In future work, it would be valuable to quantify the impact of this multiphase structure on the high energy emission from AGN powered galactic winds.

\section*{Acknowledgements} This work was supported in part by NASA ATP Grant 12-ATP12-0183 and by a Simons Investigator Award to EQ from the Simons Foundation.
  CAFG is supported by NASA through Einstein Postdoctoral Fellowship Award number PF3-140106

\bibliographystyle{mn2e}
\bibliography{paper_draft_final}
\end{document}